\documentclass{PoS}

\usepackage{amsmath}
\usepackage{amsthm} 

\newcommand{\N}{\nonumber}
\newcommand{\ep}{\varepsilon}
\newcommand\sign{{\rm sign}}
\newcommand\HA{{\rm H}}
\newcommand\Mvec{\,\mbox{\bf M}}
\newcommand{\M}[2]{\ensuremath{\mathbf{M}\left[#1\right]\left(#2\right)}}
\newcommand{\Hs}[2]{\ensuremath{\mathop{\text{H}}\nolimits^*_{#1}\left(#2\right)}}

\newtheorem{theorem}{Theorem}

\newtheorem{example}{Example}

\title{{\footnotesize DESY 14--131, DO-TH 14/16, SFB/CPP-14-55, LPN14-093}\\
Nested (inverse) binomial sums and new iterated integrals for massive Feynman diagrams}

\ShortTitle{Nested binomial sums and new iterated integrals}

\author{Jakob Ablinger$^a$, Johannes Bl\"{u}mlein$^b$, \speaker{Clemens G. Raab}%
         \thanks{This work was 
supported in part by DFG Sonderforschungsbereich Transregio 9, Computergest\"utzte Theoretische Teilchenphysik, 
the Austrian Science Fund (FWF) grants P20347-N18 and SFB F50 (F5009-N15), the European
Commission through contract PITN-GA-2010-264564 ({LHCPhenoNet}) and PITN-GA-2012-316704 ({HIGGSTOOLS}).} $^b$, 
Carsten Schneider$^a$\\
        $^a$ Research Institute for Symbolic Computation (RISC), Johannes Kepler Universit\"{a}t, Altenbergerstra{\ss}e 69, 4040 Linz, Austria\\
        $^b$ Deutsches Elektronen-Synchrotron (DESY), Platanenallee 6, 15738 Zeuthen, Germany\\
        E-mails: \email{jakob.ablinger@risc.jku.at}, \email{johannes.bluemlein@desy.de}, \email{clemens.raab@desy.de}, \email{carsten.schneider@risc.jku.at}}

\abstract{Nested sums containing binomial coefficients occur in the computation of massive operator matrix elements. Their associated iterated integrals lead to alphabets including radicals, for which we determined a suitable basis. We discuss algorithms for converting between sum and integral representations, mainly relying on the Mellin transform. To aid the conversion we worked out dedicated rewrite rules, based on which also some general patterns emerging in the process can be obtained.}

\FullConference{Loops and Legs in Quantum Field Theory -- LL2014,\\
                 27 April 2014 -- 02 May 2014\\
                 Weimar, Germany}

\begin{document}

\section{Introduction}

\noindent
In the computation of massive operator matrix elements at 3--loop order in QCD and for the  corresponding heavy flavor 
Wilson coefficients in deep-inelastic scattering the functional space spanning the Feynman-integrals is extended. While
up to 2--loop order all terms can be expressed in nested harmonic sums \cite{Vermaseren:1998uu,Blumlein:1998if} in Mellin--$N$ 
space and harmonic polylogarithms \cite{RemiddiVermaseren} in $x$-space, here also nested finite (inverse) binomial sums occur 
in a series of the Feynman diagrams \cite{Hasselhuhn12,Hasselhuhn,BinomialSumsNuclB}. Besides nested sums over the binomials $\binom{2k}{k}^{\pm 1}$ 
also generalized harmonic sums \cite{Moch:2001zr,Ablinger:2013cf} occur as standalone objects or inside of the binomial sums. The corresponding diagrams 
are characterized carrying 
two massive fermion lines of equal mass or belong to the $V$-topology in case of a single heavy mass 
\cite{BinomialSumsNuclB}.\footnote{For surveys on the function-spaces describing zero- and single scale Feynman integrals, see
Refs.~\cite{Ablinger:2013jta,Ablinger:2013eba}.}

In the following we consider finite sums of the form
\begin{equation}\label{eq:SUM1}
 \sum_{i_1=1}^Na_1(i_1)\sum_{i_2=1}^{i_1}a_2(i_2)\dots\sum_{i_k=1}^{i_{k-1}}a_k(i_k),
\end{equation}
where the summands are of the form
\begin{equation}
 a_j(N)=a(N;b_j,c_j,m_j)=\binom{2N}{N}^{b_j}\frac{c_j^N}{N^{m_j}},
\end{equation}
with 
\begin{equation}
 b_j \in \{-1,0,1\},\quad c_j \in \mathbb{R}\setminus\{0\},\quad m_j \in \mathbb{N}. 
\end{equation}
We also treat some examples with a slightly more general structure, e.g.
\begin{equation}\label{eq:SUM:3}
 a_j(N)=\frac{c_j^N}{\displaystyle(2N+1)\binom{2N}{N}}.
\end{equation}
In Ref.~\cite{BinomialSums} we have recently presented an algorithmic treatment of these sums and, associated to them, explored iterated 
integrals over an alphabet containing also square root-valued letters. The representation of the finite (inverse) binomial sums
in terms of Mellin transforms of their associated iterated integrals is instrumental for their asymptotic representation
to be derived in analytic form.
This is needed in solving the corresponding sums appearing in physical problems as well as in the 
intermediary summation steps and 
for the treatment of infinite sums in general, see also \cite{Gluza}. 
Moreover, these representations are of importance for the analytic 
continuation of theses sums into the complex plane. 
In Ref.~\cite{BinomialSums} we also developed algorithms to map iterated 
integrals over square root-valued alphabets back into the associated nested sums. In the present calculation we made extensive 
use of the packages {\tt Sigma} \cite{SIG1,SIG2}, based on advanced symbolic summation algorithms in the setting of difference 
fields \cite{Karr:81,Petkov:92,Schneider:01,Schneider:05a,Schneider:08c,Schneider:10a,Schneider:10b,Schneider:10c,
Schneider:13b}, and the packages {\tt EvaluateMultiSums}, {\tt SumProduction} \cite{Ablinger:2010pb,Blumlein:2012hg,Schneider:2013zna}, and 
{\tt HarmonicSums}
\cite{Ablinger:2010kw,Ablinger:2011te,Ablinger:2013cf,Ablinger:2013hcp}.

In this presentation we give a brief summary of our recent algorithmic results \cite{BinomialSums} treating these sums and 
new iterated integrals associated to them.
In Section~\ref{sec:IteratedInt} we introduce the corresponding iterated integrals involving square roots in their integrands 
and define a basis of them.
In Section~\ref{sec:MellinRules} we take a computational look at the Mellin transforms of iterated integrals and 
give a criterion on when they can be expressed in terms of nested (inverse) binomial sums.
The reverse task of finding representations of nested (inverse) binomial sums in terms of Mellin transforms is considered 
in Section~\ref{sec:MellinRep}, where we outline a method, which heavily relies on computing convolution integrals.
Particularly for this purpose we provide a set of rewrite rules and indicate general patterns based on them.
In Section~\ref{sec:GenFun} we consider infinite nested (inverse) binomial sums and show one method how they can be expressed in terms of iterated integrals again by rewrite rules.

\section{Iterated integrals and the Mellin transform}

\subsection{Iterated integrals}
\label{sec:IteratedInt}

\noindent
We consider iterated integrals on the interval $x \in [0,1]$. They are indexed by symbols identifying the integrands used. 
These are called letters and form words over the corresponding alphabet. In analogy to harmonic polylogarithms 
\cite{RemiddiVermaseren} we define
\begin{eqnarray}
 \Hs{\emptyset}{x} &=& 1\\
 \Hs{\sf b,\vec{c}}{x} &=& \int_x^1dt\,b(t)\Hs{\sf \vec{c}}{t}.
\end{eqnarray}
Note that integration is over the interval $[x,1]$ in contrast to harmonic polylogarithms $\mathrm{H}_{\sf \vec{a}}(x)$, where integration is over $[0,x]$. We use the star to make notation unambiguous.

In general all iterated integrals satisfy the shuffle relations, see Ref.~\cite{Ree}, the ones we consider are no exception.
It is convenient to consider alphabets such that all algebraic relations among the iterated integrals over the given alphabet are already induced by the shuffle relations and there are no additional algebraic relations.
For integrands with root-singularities this property can be ensured by the choice of the alphabet stated below.
This relies on a theorem proven in Ref.~\cite{DeneufchatelEtAl}, which gives a criterion on the linear independence of iterated integrals over a given alphabet.
Since any polynomial expression in terms of iterated integrals can be reduced to a linear combination of iterated integrals over the same alphabet by shuffling, establishing linear independence implies that all algebraic relations among the iterated integrals are due to shuffling.
Using results from Refs.~\cite{Risch,Trager79} the iterated integrals over the alphabet 
defined below can be proven to be linearly independent over the algebraic functions
\begin{eqnarray}
 f_a(x)&:=&\frac{\sign(1-a-0)}{x-a},\\
 f_{\{a_1,\dots,a_k\}}(x)&:=&f_{a_1}(x)^{1/2}\dots f_{a_k}(x)^{1/2}\quad\quad k\ge2,\\
 f_{(a_0,\{a_1,\dots,a_k\})}(x)&:=&f_{a_0}(x)f_{a_1}(x)^{1/2}\dots f_{a_k}(x)^{1/2}\quad\quad k\ge1,\\
 f_{(\{a_1,\dots,a_k\},j)}(x)&:=&x^jf_{\{a_1,\dots,a_k\}}(x) \quad\quad j\in\{1,\dots,k-2\}.
\end{eqnarray}
In our applications it is actually sufficient to allow at most two root-singularities, hence we can restrict to the following 
cases:
\begin{eqnarray}
 f_a(x)&:=&\frac{\sign(1-a-0)}{x-a},\\
 f_{(a,\{b\})}(x)&:=&f_{a}(x)\sqrt{f_{b}(x)},\\
 f_{\{a,b\}}(x)&:=&\sqrt{f_{a}(x)}\sqrt{f_{b}(x)},\label{eq:LetterSqrtSqrt}\\
 f_{(a,\{b,c\})}(x)&:=&f_{a}(x)\sqrt{f_{b}(x)}\sqrt{f_{c}(x)}.
\end{eqnarray}
Already in 2004 the following six letters with root-singularities were considered in the context of 2-loop integrals with massive propagators \cite{AgliettiBonciani}:
\begin{equation}
 \frac{1}{\sqrt{x(4 \pm x)}} \quad\text{and}\quad \frac{1}{(1 \pm x)\sqrt{x(4 \pm x)}}.
\end{equation}

\subsection{Mellin transforms of iterated integrals}
\label{sec:MellinRules}

\noindent
There are several general methods to compute Mellin transforms 
\begin{eqnarray}
\Mvec[f(x)](N) = \int_0^1~dx~x^N~f(x)
\end{eqnarray}
for a large class of so-called D-finite functions, which are described in Ref.~\cite{BinomialSums}.
We do not discuss these methods here.
Instead we put emphasis on more specialized results, which provide rewrite rules to directly compute the Mellin transform of certain generalized harmonic polylogarithms that involve square roots as introduced above.
These iterated integrals are special cases of D-finite functions and the corresponding rewrite rules enable us to obtain the result in terms of nested (inverse) binomial sums.
In order to do so we proceed recursively by applying the formulae given below.
They mainly work by reducing the depth of the iterated integrals occurring inside the Mellin transform by reducing the Mellin transform involving $f(x)$ to one involving $f^\prime(x)$.

It is well known that the Mellin transform satisfies the following identities.
\begin{eqnarray}
 \M{f(x)}{N} &=& \frac{1}{N+1}\left(f(1)-\M{xf^\prime(x)}{N}\right)\\*
 \M{\frac{f(x)}{x-c}}{N} &=& c^N\left(\int_0^1dx\frac{f(x)}{x-c} \,+\, \sum_{i=1}^N\frac{1}{c^i}\M{f(x)}{i-1}\right).
\end{eqnarray}
Furthermore, we were able to obtain identities for certain cases where the input involves square roots. Finding these identities involved {\tt Singular} \cite{Singular} and {\tt HolonomicFunctions} \cite{HF}.
Let $a \in \mathbb{C}\setminus[0,\infty[$, then for all $N \in \mathbb{N}$ and sufficiently regular $f(x)$ we have
\begin{eqnarray}
 \M{\frac{f(x)}{\sqrt{x-a}}}{N} &=& \frac{(4a)^N}{(2N+1)\binom{2N}{N}}\Bigg(\int_0^1dx\frac{f(x)}{\sqrt{x-a}}\\*
 && +2\sqrt{1-a}f(1)\sum_{i=1}^N\frac{\binom{2i}{i}}{(4a)^i}-2\sum_{i=1}^N\frac{\binom{2i}{i}}{(4a)^i}
\M{\sqrt{x-a}f^\prime(x)}{i}\Bigg)
\nonumber\\
\M{\frac{f(x)}{\sqrt{x(x-a)}}}{N} &=& \left(\frac{a}{4}\right)^N\binom{2N}{N}\Bigg(\int_0^1dx\frac{f(x)}{\sqrt{x(x-a)}}
\\*
 && +\sqrt{1-a}f(1)\sum_{i=1}^N\frac{(4/a)^i}{i\binom{2i}{i}}-\sum_{i=1}^N\frac{(4/a)^i}
{i\binom{2i}{i}}\M{\sqrt{\frac{x-a}{x}}f^\prime(x)}{i}\Bigg)
\nonumber\\
\M{\sqrt{\frac{x}{x-a}}f(x)}{N} &=& \frac{1}{2}\left(\frac{a}{4}\right)^N\binom{2(N+1)}{N+1}\Bigg(\int_0^1dx
\sqrt{\frac{x}{x-a}}f(x)
 +\sqrt{1-a}f(1)\sum_{i=1}^N\frac{(4/a)^i}{(2i+1)\binom{2i}{i}} \nonumber\\
&& -\sum_{i=1}^N\frac{(4/a)^i}{(2i+1)\binom{2i}{i}}\M{\sqrt{x(x-a)}f^\prime(x)}{i}\Bigg).
\end{eqnarray}

Moreover, based on these formulae we can give some sufficient conditions on when the Mellin transform can be written in terms of nested (inverse) binomial sums, which is summarized in the theorem below.
Note that analytic continuation can be used to relax some of the restrictions on the position of the singularities.
\begin{theorem}
 Let $r_0(x),\dots,r_k(x) \in \mathbb{C}(x)\setminus\{0\}$ without a pole at $x=1$,
 let $p_0(x),\dots,p_k(x) \in \mathbb{C}[x]\setminus\{0\}$ with all their roots in $]{-}\infty,0]$,
 and set $h_i(x):=\frac{r_i(x)}{\sqrt{p_i(x)}}$.
 Assume that each of the products $h_0(x)\cdots h_i(x)$ is of one of the forms $r(x)$, $\frac{r(x)}{\sqrt{x}}$, $\frac{r(x)}{\sqrt{x-a}}$, or $\frac{r(x)}{\sqrt{x(x-a)}}$ for some $r(x)\in\mathbb C(x)\setminus\{0\}$ (not necessarily all of the same form).
 If the integral $\int_0^1dx\,h_0(x)\HA^*_{\sf h_1,\dots,h_k}(x)$ exists, then the Mellin transform $\M{h_0(x)\HA^*_{\sf h_1,\dots,h_k}(x)}{N}$ is expressible in terms of nested (inverse) binomial sums.
\end{theorem}

\section{Mellin representations of finite binomial sums}
\label{sec:MellinRep}

\noindent
Our aim is to represent nested (inverse) binomial sums in terms of a number of Mellin transforms each weighted by an exponential term.
More precisely, we aim at representations of the form
\begin{equation}\label{eq:IntRepForm}
c_0+\sum_{j=1}^k c_j^N\M{f_j(x)}{N},
\end{equation}
where the constants $c_j$ and functions $f_j(x)$ do not depend on $N$.
Such representations can be computed merely by relying on the properties of the Mellin transform 
combined with additional tools for computing Mellin convolutions discussed below.
Due to the use of the summation property Eq.~(2.3) in \cite{BinomialSums} often it is not necessary to 
specify the 
constant $c_0$ separately because often we have
\begin{equation}
 c_0 = -\M{\sum_{j=1}^kf_j(x)}{0}
\end{equation}
and then it is also possible to write \eqref{eq:IntRepForm} as
\begin{equation}\label{eq:IntRepForm2}
\sum_{j=1}^k \int_0^1dx \frac{(c_jx)^N-1}{x-\frac{1}{c_j}}g_j(x).
\end{equation}

We base the calculations of Mellin representations on the following basic ones, which are all we need as a starting point:
\begin{eqnarray}
 \frac{1}{N} &=& \M{\frac{1}{x}}{N}\label{eq:RepresentInverse}\\
 \binom{2N}{N} &=& \frac{4^N}{\pi}\M{\frac{1}{\sqrt{x(1-x)}}}{N}\label{eq:RepresentBinomial}\\
 \frac{1}{N\displaystyle\binom{2N}{N}} &=& \frac{1}{4^N}\M{\frac{1}{x\sqrt{1-x}}}{N}\label{eq:RepresentInverseBinomial}.
\end{eqnarray}
From these we can obtain integral representations for sums and nested sums step by step. In general the computation proceeds as follows. Starting from the innermost sum we move outwards maintaining an integral representation of the subexpressions visited so far. For each intermediate sum
\begin{equation}\label{eq:AlgorithmIntermediate}
\sum_{i_j=1}^Na_j(i_j)\sum_{i_{j+1}=1}^{i_j}a_{j+1}(i_{j+1})\dots\sum_{i_k=1}^{i_{k-1}}a_k(i_k)
\end{equation}
this first involves setting up an integral representation for the summand $a_j(N)$ of the form \eqref{eq:IntRepForm}. This may require computation of Mellin convolutions, which we will describe in more detail below. Next we obtain an integral representation of the same form of
\begin{equation}
a_j(N)\sum_{i_{j+1}=1}^Na_{j+1}(i_{j+1})\dots\sum_{i_k=1}^{i_{k-1}}a_k(i_k)
\end{equation}
by Mellin convolution with the result for the inner sums computed so far. Then by the summation property 
we obtain an integral representation for the sum \eqref{eq:AlgorithmIntermediate}. These steps are repeated until the outermost sum has been processed.

\subsection{Convolution integrals}

\noindent
As mentioned above Mellin convolutions need to be computed at several points in our computation of Mellin representations.
Convolution integrals are the most challenging part of the computation, so we take a closer look on how we can compute them.
The convolution formula 
\begin{eqnarray}
A(x) \otimes B(x) = \int_0^1 dx_1 \int_0^1 dx_2 \delta(x-x_1x_2) A(x_1) B(x_2)   = \int_x^1 \frac{dt}{t} A(t) B\left(\tfrac{x}{t}\right)
\end{eqnarray}
gives us a definite integral depending on a continuous parameter and hence can be written in the form
\begin{equation}\label{eq:ParameterIntegral}
 F(x) = \int_{a(x)}^{b(x)}dt\,f(x,t).
\end{equation}
One general strategy to obtain a closed form for such integrals is to first set up a differential equation satisfied by $F(x)$ and then obtain a solution of this equation satisfying appropriate initial conditions.
In the first step we exploit the principle of differentiation under the integral.
If we have a relation for the integrand $f(x,t)$ of the form
\begin{equation}\label{eq:ParametricIntegration}
 c_m(x)\frac{\partial^mf}{\partial{x}^m}(x,t)+\dots+c_0(x)f(x,t) = \frac{\partial g}{\partial{t}}(x,t)
\end{equation}
for some coefficients $c_i(x)$ independent of $t$ and some function $g(x,t)$, then by applying $\int_x^1dt$ this gives rise to 
a linear ordinary differential equation for the integral $F(x)$
\begin{equation}\label{eq:ResultingODE}
 c_m(x)F^{(m)}(x)+\dots+c_0(x)F(x) = g(x,b(x))-g(x,a(x))+\text{additional boundary terms}.
\end{equation}
In the presence of singularities proper care has to be taken when evaluating the right hand side of this relation.
There are several computer algebra algorithms for different types of integrands $f(x,t)$ which, given $f(x,t)$, compute relations of the form \eqref{eq:ParametricIntegration}.
They either utilize differential fields \cite{Risch,SingerSaundersCaviness,Bronstein,RaabPhD} or holonomic systems and Ore algebras \cite{AlmkvistZeilberger,Chyzak,Koutschan,ChenKauersKoutschan}.
After obtaining a differential equation for $F(x)$ we need to solve it explicitly, preferably in terms of iterated integrals.
If the differential equation factors completely into first-order factors with rational function coefficients, then this can be achieved.
In practice all the specific sums from Ref.~\cite{BinomialSumsNuclB} we considered give rise to differential equations with this property.
Even more is true, the first-order factors all have algebraic functions of degree at most two as their solutions, hence the solutions of the differential equations are of the form
\begin{equation}
 \frac{r_1(x)}{\sqrt{p_1(x)}} \int dx \frac{r_2(x)}{\sqrt{p_2(x)}} \int dx \dots \int dx \frac{r_k(x)}{\sqrt{p_k(x)}},
\end{equation}
where $r_i(x)$ are rational functions and $p_i(x)$ are square-free polynomials.
Using a dedicated rewrite procedure based on integration by parts we can write a basis of the solution space in terms of the functions $\mathrm{H}^*$ over the alphabet defined earlier, which is then used to match initial conditions.
Already Hermite considered a reduction procedure for simple integrals of the form $\int dx \frac{r(x)}{\sqrt{p(x)}}$ similar to ours, see Ref.~\cite{Hermite83}.

Instead of relying just on the form \eqref{eq:ParameterIntegral}, alternatively one can also try to exploit the special structure present in the convolution integrals of iterated integrals.
It turns out that for most convolutions needed in the context of finding Mellin representations for nested (inverse) binomial sums one of the factors is just an algebraic function and the other involves iterated integrals.
Our approach was inspired by Refs.~\cite{GeddesLeLi,ChenKauersSinger,BostanEtAl} where the respective authors proposed algorithms which, for certain types of integrands, compute the differential operators in \eqref{eq:ParametricIntegration} already in partially factored form.
In analogy to the formulae used to compute Mellin transforms of iterated integrals, the general idea here is to recursively reduce the convolution of iterated integrals to convolution integrals where the depth of the iterated integrals has been reduced.
In the process the original convolution integral is directly converted into iterated integrals without computing intermediate differential equations.
Again this is done by specialized rewrite rules, of which we list only a few here.
A more general list of formulae that we found is included in Ref.~\cite{BinomialSums}.
\begin{eqnarray}
 \int_x^1dt\frac{f(t)}{(t-c)\sqrt{t-x}} &=& \frac{1}{\sqrt{x-c}}\int_x^1dt\frac{1}{\sqrt{t-c}}\bigg(\frac{f(1)}{\sqrt{1-t}}-\int_t^1du\frac{f^\prime(u)}{\sqrt{u-t}}\bigg)\\
 \int_x^1dt\frac{f(t)}{\sqrt{t-a}\sqrt{t-x}} &=& \int_x^1dt\frac{1}{t-a}\bigg(\frac{\sqrt{1-a}f(1)}{\sqrt{1-t}}-\int_t^1du\frac{\sqrt{u-a}f^\prime(u)}{\sqrt{u-t}}\bigg)\\
 \int_x^1dt\frac{\sqrt{t}f(t)}{(t-c)\sqrt{t-x}} &=& \int_x^1dt\frac{1}{t}\bigg(\frac{f(1)}{\sqrt{1-t}}-\int_t^1du\frac{\sqrt{u}f^\prime(u)}{\sqrt{u-t}}\bigg)+\\*
 &&+\frac{c}{\sqrt{x-c}}\int_x^1dt\frac{1}{t\sqrt{t-c}}\bigg(\frac{f(1)}{\sqrt{1-t}}-\int_t^1du\frac{\sqrt{u}f^\prime(u)}{\sqrt{u-t}}\bigg)\nonumber
\end{eqnarray}

In the Mellin representations we computed, see Ref.~\cite{BinomialSums}, some patterns emerge. They can be proven based on the 
rewrite rules. These patterns show how absorbing simple pre-factors into the Mellin transform changes the iterated integrals 
occurring there. Here we list a few of them and refer to Ref.~\cite{BinomialSums} for a more comprehensive list. 
Let 
$a_0,\dots,a_k,c<0$ and to shorten notation we let ${\bf b_i}=\{a_i,a_{i+1}\}$ for $i \in \{0,\dots,k-1\}$ and ${\bf b_k}=\{1,a_k\}$.
\begin{eqnarray}
 \binom{2N}{N}\M{\frac{\Hs{a_1,\dots,a_k}{x}}{x-a_0}}{N} &=& \frac{4^N}{\pi}\M{\frac{\Hs{{\bf b_0},\dots,{\bf b_k}}{x}}{\sqrt{x(x-a_0)}}}{N}\\
  \frac{1}{(2N+1)\binom{2N}{N}}\M{\frac{x\Hs{{\bf b_0},\dots,{\bf b_k}}{x}}{(x-c)\sqrt{x(x-a_0)}}}{N} &=& 
\frac{\pi}{2{\cdot}4^N}\M{\frac{\Hs{(a_0,\{c\}),a_1,\dots,a_k}{x}}{\sqrt{x-c}}}{N}.
\end{eqnarray}

We finally give an example for the Mellin representation of a more involved sum, see \cite{BinomialSumsNuclB},
\begin{eqnarray}
\lefteqn{\sum_{i=1}^N \frac{1}{\displaystyle (i+1) \binom{2i}{i}} \sum_{j=1}^i
\binom{2j}{j} \frac{1}{j}
S_{-2}(j)
\ =\
\int_0^1 dx \frac{(-x)^N-1}{x+1}\Biggl(\frac{x}{\sqrt{x+\frac{1}{4}}}\HA^*_{\sf w_{14},-1,0}(x)
}\N\\*&&
-\frac{x}{2}\HA^*_{\sf w_{14},w_{14},-1,0}(x)\Biggr)
-\frac{\zeta_2}{2}\int_0^1 dx \frac{x^N-1}{x-1}\left(\frac{x}{\sqrt{x-\frac{1}{4}}}\HA^*_{\sf w_8}(x)-\frac{x}{2}\HA^*_{\sf 
w_8,w_8}(x)\right)
\N\\*&&
-\frac{\HA^*_{\sf -\frac{1}{4},w_2,w_2,w_1}(0)}{\pi}\int_0^1 dx 
\frac{(\frac{x}{4})^N-1}{x-4}\left(\frac{x}{\sqrt{1-x}}-\frac{x}{2}\HA^*_{\sf w_3}(x)\right).
\end{eqnarray}
This iterated integral is based on the square root-valued letters 
\begin{eqnarray}
f_{\sf w_1} &=& \frac{1}{\sqrt{x}\sqrt{1-x}} \\
f_{\sf w_2} &=& \frac{1}{\sqrt{x}\sqrt{1+x}} \\
f_{\sf w_3} &=& \frac{1}{x\sqrt{1-x}}\\
f_{\sf w_8} &=& \frac{1}{x \sqrt{x-\tfrac{1}{4}}}\\
f_{\sf w_{14}} &=& \frac{1}{x\sqrt{x+\tfrac{1}{4}}}~.
\end{eqnarray}

\section{Representing infinite binomial sums in terms of iterated integrals}
\label{sec:GenFun}

\noindent
We also consider power series whose coefficients are given by nested (inverse) binomial sums. Infinite (inverse) binomial sums were considered in several contexts already, see Refs.~\cite{FleischerEtAl,JegerlehnerKalmykovVeretin,DavydychevKalmykov,Weinzierl,KalmykovWardYost} where related results were obtained.
Previous results often obtain different integral representations, which are not explicitly expressed in terms of iterated integrals.
We aim at a direct way of writing the infinite sums in terms of iterated integrals.
In analogy to the formulae we developed in the previous sections for the computation of Mellin transforms and convolution integrals, we find rewrite rules which can be applied recursively in order to express infinite nested (inverse) binomial sums in terms of iterated integrals.
First we summarize a few well known properties.
\begin{eqnarray}
 \sum_{n=1}^\infty\frac{x^n}{n}f(n) &=& \int_0^xdt\frac{1}{t}\sum_{n=1}^\infty t^nf(n)\label{eq:GenFunReciprocal}\\
 \sum_{n=1}^\infty x^n\sum_{i=1}^nf(i) &=& \frac{1}{1-x}\sum_{n=1}^\infty x^nf(n)\\
 \sum_{n=1}^\infty\frac{x^n}{n+1}f(n) &=& \frac{1}{x}\int_0^xdt\sum_{n=1}^\infty t^nf(n)\\*
 &=& \sum_{n=1}^\infty\frac{x^n}{n}f(n)-\frac{1}{x}\int_0^xdt\sum_{n=1}^\infty \frac{t^n}{n}f(n)
\end{eqnarray}
Next, we give some identities specifically designed for expressions involving binomial coefficients.
\begin{eqnarray}
 \sum_{n=1}^\infty x^n\binom{2n}{n}\sum_{i=1}^nf(i) &=& \frac{1}{4\sqrt{\frac{1}{4}-x}}\int_0^xdt\frac{1}{t\sqrt{\frac{1}{4}-t}}\sum_{n=1}^\infty t^nn\binom{2n}{n}f(n)\label{eq:GenFunBinomialSum}\\
 \sum_{n=1}^\infty \frac{x^n}{n\binom{2n}{n}}\sum_{i=1}^nf(i) &=& \frac{\sqrt{x}}{\sqrt{4-x}}\int_0^xdt\frac{1}{\sqrt{t}\sqrt{4-t}}\sum_{n=0}^\infty \frac{t^n}{\binom{2n}{n}}f(n+1)\label{eq:GenFunInverseBinomialSum1a}\\*
  &=& \sum_{n=1}^\infty \frac{x^n}{n\binom{2n}{n}}f(n)+\frac{\sqrt{x}}{\sqrt{4-x}}\int_0^xdt\frac{1}{\sqrt{t}\sqrt{4-t}}\sum_{n=1}^\infty \frac{t^n}{\binom{2n}{n}}f(n)\label{eq:GenFunInverseBinomialSum1b}\\
 \sum_{n=1}^\infty\frac{x^n}{(2n+1)\binom{2n}{n}}\sum_{i=1}^nf(i) &=& \frac{2}{\sqrt{x}\sqrt{4-x}}\int_0^xdt
\frac{1}{\sqrt{t}\sqrt{4-t}}\sum_{n=1}^\infty\frac{t^n}{\binom{2n}{n}}f(n).
\end{eqnarray}
Note that for $f(n):=\delta_{n,1}$ we obtain the following identities as special cases from the formulae above
\begin{eqnarray}
 \sum_{n=1}^\infty x^n\binom{2n}{n} &=& \frac{1}{2\sqrt{\frac{1}{4}-x}}-1\label{eq:GenFunBinomial}\\
 \sum_{n=1}^\infty \frac{x^n}{n\binom{2n}{n}} &=& \frac{\sqrt{x}}{\sqrt{4-x}}\int_0^xdt\frac{1}{\sqrt{t}\sqrt{4-t}}\label{eq:GenFunInverseBinomial1}\\
 \sum_{n=1}^\infty\frac{x^n}{(2n+1)\binom{2n}{n}} &=& \frac{2}{\sqrt{x}\sqrt{4-x}}\int_0^xdt\frac{1}{\sqrt{t}
\sqrt{4-t}}-1.\label{eq:GenFunInverseBinomial2}
\end{eqnarray}
These rewrite rules have also been applied recently in Ref.~\cite{Gluza}. Note, however, that their applicability is not limited to sums of low depth.
\begin{example}
 For illustration of the use of the formulae above consider the simple infinite inverse binomial sum
 \begin{equation}\label{eq:GenFunExample1}
  \sum_{n=1}^\infty\frac{x^n}{n\binom{2n}{n}}S_2(n).
 \end{equation}
 Applying \eqref{eq:GenFunInverseBinomialSum1b} and then \eqref{eq:GenFunReciprocal} we obtain
 \begin{eqnarray*}
  \sum_{n=1}^\infty\frac{x^n}{n\binom{2n}{n}}S_2(n) &=& \sum_{n=1}^\infty \frac{x^n}{n^3\binom{2n}{n}}+\frac{\sqrt{x}}{\sqrt{4-x}}\int_0^xdt\frac{1}{\sqrt{t}\sqrt{4-t}}\sum_{n=1}^\infty \frac{t^n}{n^2\binom{2n}{n}}\\
  &=&\int_0^xdt\frac{1}{t}\int_0^tdu\frac{1}{u}\sum_{n=1}^\infty\frac{u^n}{n\binom{2n}{n}}+\frac{\sqrt{x}}{\sqrt{4-x}}\int_0^xdt\frac{1}{\sqrt{t}\sqrt{4-t}}\int_0^tdu\frac{1}{u}\sum_{n=1}^\infty\frac{u^n}{n\binom{2n}{n}}.
 \end{eqnarray*}
 Now, by virtue of \eqref{eq:GenFunInverseBinomial1} we obtain the result
 \begin{eqnarray}
  \sum_{n=1}^\infty\frac{x^n}{n\binom{2n}{n}}S_2(n) &=& \int_0^xdt\frac{1}{t}\int_0^tdu\frac{1}{\sqrt{u}\sqrt{4-u}}\int_0^udv\frac{1}{\sqrt{v}\sqrt{4-v}}
\N\\&&
  +\frac{\sqrt{x}}{\sqrt{4-x}}\int_0^xdt\frac{1}{\sqrt{t}\sqrt{4-t}}\int_0^tdu\frac{1}{\sqrt{u}\sqrt{4-u}}\int_0^udv\frac{1}{\sqrt{v}\sqrt{4-v}}\label{eq:GenFunExample1Result}\\
  &=&\mathrm{H}_{0,\{0,4\},\{0,4\}}(x)+\frac{\sqrt{x}}{\sqrt{4-x}}\mathrm{H}_{\{0,4\},\{0,4\},\{0,4\}}(x).
 \end{eqnarray}
 Recall the notation \eqref{eq:LetterSqrtSqrt} and $\mathrm{H}_{\vec{a}}(x)$, which refers to integration over the interval $[0,x]$.

 Finally, in the present case we may also apply the change of variable $x=-\frac{(1-y)^2}{y}$, cf.~\cite{DavydychevKalmykov}, in the iterated integrals to remove the square roots in their integrands. The result can then be expressed in terms of polylogarithms:
 \begin{equation}
  \sum_{n=1}^\infty\frac{x^n}{n\binom{2n}{n}}S_2(n) = 
2\text{\rm Li}_3(y)-2\ln(y)\text{\rm Li}_2(y)-\ln(y)^2\ln(1-y)+\frac{y\ln(y)^3}{3(1+y)}-2\zeta(3).
 \end{equation}
\end{example}

\section{Conclusions}

\noindent
We studied finite nested (inverse) binomial sums and their associated iterated integrals which emerge in massive 3-loop 
calculations in QCD. The sums and the iterated integrals are related by the Mellin transform. The iterated integrals are built
over alphabets containing the letters which form the harmonic polylogarithms and their generalization \cite{Ablinger:2013cf} 
and a large number of square root-valued letters. Algorithms were presented to transform a given nested (inverse) binomial sum
into its associated iterated integral. In Ref.~\cite{BinomialSums} we also developed methods for 
the reverse way, one of which was outlined here.
The present algorithms are needed to express the corresponding physics results both in $N$- and in $x$-space and to obtain
the asymptotic series of sums, being required in various summation problems.  
We briefly commented also on infinite (inverse) binomial sums, also emerging in many physical 
problems, e.g. as generating functions.


\end{document}